\documentclass[12pt,preprint]{aastex}

\begin{document}

\title{The Redshifted Excess in Quasar C IV Broad Emission Lines}
\author{Brian Punsly\altaffilmark{1}} \altaffiltext{1}{4014 Emerald Street No.116, Torrance CA,
USA 90503 and ICRANet, Piazza della Repubblica 10 Pescara 65100,
Italy, brian.punsly@verizon.net or brian.punsly@comdev-usa.com}

\begin{abstract}
In this Letter, the Evans and Koratkar Atlas of Hubble Space
Telescope Faint Object Spectrograph Spectra of Active Galactic
Nuclei and Quasars is used to study the redward asymmetry in CIV
broad emission lines (BELs). It is concluded that there is a highly
significant correlation between the spectral index from 10 GHz to
1350 $\AA$ and the amount of excess luminosity in the red wing of
the CIV BEL ($>99.9999\%$ significance level for the full sample and
the radio loud subsample independently, but no correlation is found for
the radio quiet subsample). This is interpreted
as a correlation between radio core dominance and the strength of
the CIV redward asymmetry. The data implies that within the quasar
environment there is BEL gas with moderately blueshifted emission
associated with the purely radio quiet quasar phenomenon (the
accretion disk) and the radio jet emission mechanism is associated
with a redward BEL component that is most prominent for lines of
sight along the jet axis. Thus, radio quiet quasars have CIV BELs
that tend to show blueshifted excess and radio loud quasars show
either a red or blue excess with the tendency for a dominant red
excess increasing as the line of sight approaches the jet axis.
\end{abstract}


\section{Introduction}About 10\% of quasars possess powerful
relativistic radio jets (known generically as radio loud quasars
RLQs). It is not clear if there are any differences between the
accretion states of the central engines (accretion flow plus central
supermassive black hole) in RLQs and radio quiet quasars (RQQs) that
are defined by "weak" jet power.\footnote{The radio loudness, $R$,
is usually defined as a 5 GHz flux density at least 10 times larger
than the $4400 \AA$ flux density, $R=S_{5 \mathrm{GHz}}/S_{4400
\AA}>10$ with $R>10$ being a crude indication of a powerful radio
jet \citep{kel89}.} The signature of the quasar phenomenon is the
thermal continuum luminosity created by dissipation in the accretion
flow that results in a large blue/UV excess in the spectrum
\citep{sun89}. The most prominent feature of quasar optical/UV
spectra are the conspicuous broad emission lines (BELs). To first
order, the continuum and BELs in RLQS and RQQs are remarkably
similar \citep{cor94,zhe97,tel02}. Systematic differences in the two
families of spectra are only revealed by the study of subtle lower
order spectral features \citep{cor98}. The differences are so small
that it is not clear if these features result from a difference in
environment, or directly from emission induced (or suppressed) by
the jet proper or actually a difference related to the accretion
flow. In this paper, we concentrate on the properties of the CIV BEL
in hopes of shedding light on the connection between radio jet
propagation and the accretion state. The asymmetry of the CIV BEL
has received significant attention in the past. Most efforts have
been concentrated on the blueshift seen in predominantly radio quiet
samples (see \citet{wil84,bro94,mar96,bas05,sul07} and references
therein). There is also a more limited discussion of redward
asymmetry in radio loud quasars
\citep{cor91,cor96,mar96,wil95,bac04}. Not only is this topic of
fundamental interest to the study of the quasar central engines, but
more recently it has become a point of controversy in the field of
central black hole mass estimates based on the assumption of the BEL
gas being virialized in a gravitational potential \citep{bas05}. For
high redshift sources, CIV is one of the available BELs that can be
used to estimate these virialized motions \citep{ves06}. So an
important question is whether the gas that creates the emission in
the asymmetric broad redwings represents virialized gas or some
other gas flow that is driven by other forces.
\par This study is motivated by a few anecdotal comments indicating that very large redward asymmetries
are found in the CIV BEL of some highly superluminal blazars
\citep{mar96,wil95,net95,cor97}. It is of profound physical
importance to determine whether these quasars are bizarre outliers
or an extension of a trend within the quasar population. If they are
an extension of a trend within the parent population then they are
crucial evidence in the search for the source of the redward
asymmetry. We explore this aspect through a large dataset, the HST
spectra from the Evans and Koratkar Atlas of Hubble Space Telescope
Faint Object Spectrograph Spectra of Active Galactic Nuclei and
Quasars \citet{eva04}.
\par The history of the topic of the red and blue asymmetries in the CIV BEL of
quasars is varied. Here we list some of the known results.
\begin{enumerate}
\item In \citet{cor91,cor94} a high redshift sample was used to show the
CIV BELs in steep spectrum radio loud quasars (SSQs) were more
redward asymmetric than those in flat spectrum radio loud quasars
(FSQs).
\item In \citet{wil95} there was a brief comment that core
dominated quasars (CDQs) in a small sample of RLQs have larger CIV
redward asymmetries, contrary to result 1.
\item In \citet{cor92,cor96} it was found that the CIV redward asymmetry
in quasars was correlated with the UV continuum luminosity.
\item In \citet{cor97} it was noted that most redward asymmetric
objects are RLQs and he notes that the correlation with luminosity
seen in \citet{cor96} might be a false correlation that arises
because the highest luminosity objects in that sample tended to be
radio loud.
\item Based on composite spectra from a large sample of SDSS spectra, \citet{ric02} showed that RLQs have more
redward emission in the CIV BEL relative to line center than RQQ quasars.
\item Using archived HST spectra, \citet{bac04} showed that the composite CIV spectrum of RLQs
showed a red excess relative the RQQ composite spectrum.
\item It was mentioned in \citet{mar96} that highly superluminal
blazars (apparent velocity $\sim$ 10c) showed redward asymmetry in
the CIV BEL profile.
\item In \citet{cor96} it was found that the CIV properties of RLQs
showed a larger statistical spread than those for RQQs, including
the asymmetry.
\item In \citet{bro94,wil84,mar96,bas05,cor94} a blueward asymmetry was found in samples dominated by RQQs.
In \citet{bas05}, it was determined that the blueward asymmetry of
their sample did not seem to correlate with any other spectral
properties.
\end{enumerate}
In this letter, based on an analysis of a large sample of HST spectra, a synthesis of most of the
disparate claims noted above is achieved.
\section{The HST Sample} A sample of 886 spectra from 221 sources are catalogued
in Evans and Koratkar Atlas of Hubble Space Telescope Faint Object
Spectrograph Spectra of Active Galactic Nuclei and Quasars
\citep{eva04}. This sample was used to measure the redward asymmetry
in CIV BELs. The formula that was used by \citet{wil95} was chosen
to quantify this asymmetry, $A_{25-80}$, since we want to make
contact with their subsample of "cleaner" HST spectra that were
produced from meticulous calibration and careful continuum and line
fitting. The reader is referred \citet{wil95} in order to find the
details of how the raw data was processed.
\begin{figure}
\includegraphics[width=170 mm, angle=0]{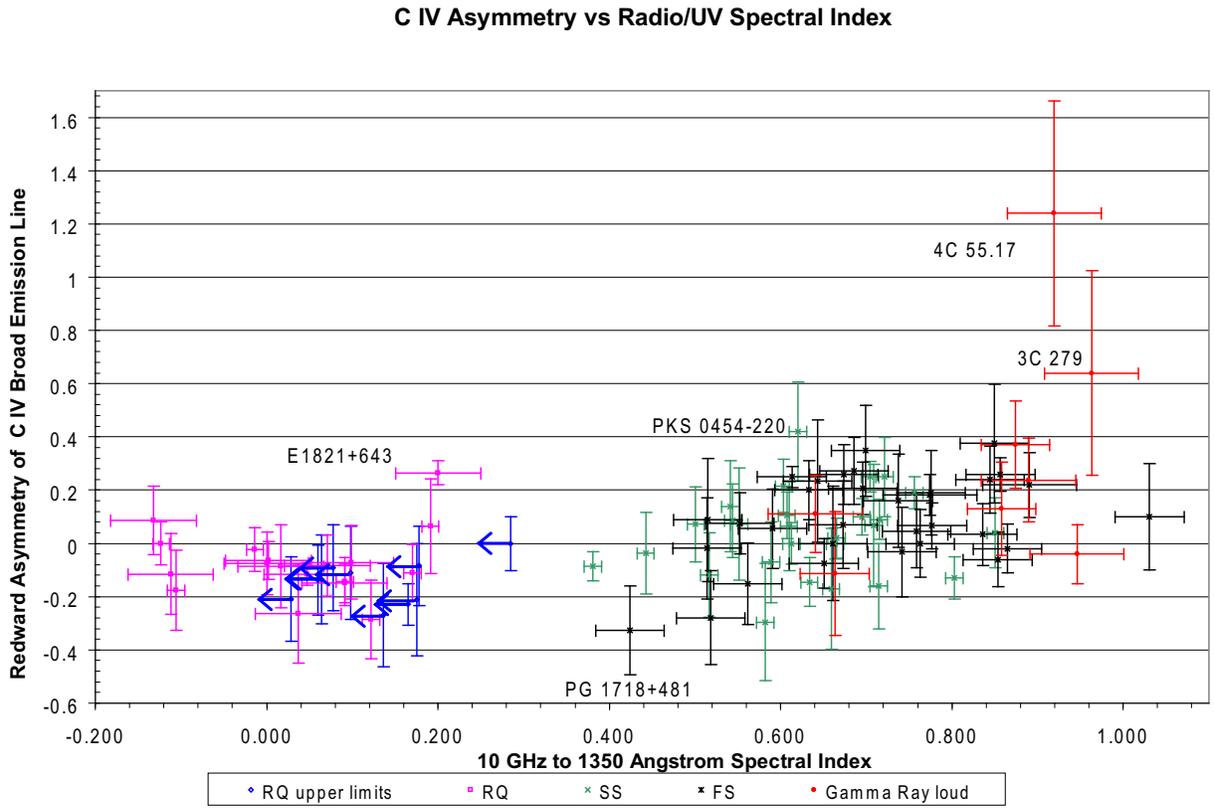}
\caption{Scatter plot of $A_{25-80}$ vs $\alpha_{10}^{UV}$. There is
a strong correlation within the total sample and also within the
radio loud subsample.}
\end{figure}
The quantity, $A_{25-80}$, is defined in terms of the full width
half maximum, FWHM, in $\AA$, the midpoint of an imaginary line
connecting a point defined at 1/4 of the peak flux density of the
BEL on the red side of the BEL to 1/4 of the peak flux density on
the blue side of the BEL, $\lambda_{25}$, and a similar midpoint
defined at 8/10 of the flux density maximum,  $\lambda_{80}$, as
\begin{equation}
A_{25-80} = \frac{\lambda_{25} - \lambda_{80}}{FWHM} \;.
\end{equation}
A positive value of $A_{25-80}$ means that there is excess flux in
the red broad wing of the BEL. A negative value of $A_{25-80}$ indicates a blueward
asymmetry of the BEL. In order to get a reliable measure of
$A_{25-80}$, one needs sufficient signal to noise to make a
meaningful measurement of $\lambda_{25}$. Also, there must not be
deep intrinsic absorption features located in the BEL that inhibit
the measurement of any of the quantities in equation (1). The number
of broad line objects in the atlas with an observed CIV profile that
also meet our minimum standards required to measure $A_{25-80}$ is
only 95 out of the 221 sources in \citet{eva04}.
\par We used the HST spectral data downloaded from MAST except in the cases where higher quality spectra
already existed in \citet{wil95}. The continuum level was set with a
local power law fit between $\sim$ 1470 $\AA$ and $\sim$ 1620 $\AA$.
Figure 1 is a plot of $A_{25-80}$ versus the spectral index from 10
GHz to 1350 $\AA$ in the quasar rest frame, $\alpha_{10}^{UV}$
(where, the flux density two point spectral index is defined by the
convention $F_{\nu} \sim \nu^{-\alpha}$). The luminosity at 1350
$\AA$ is always available when CIV is being observed and it is a
common measure of the UV continuum near the peak of the spectral
energy distribution that is utilized in virial black hole mass
estimates \citep{ves06}. The radio data is compiled from the NASA
Extragalactic Database, the FIRST and NVSS 1.4 GHz surveys and the
GB and PMNJ 5 GHz surveys. The physical interpretation of the two
point spectral index is closely related to the logarithmic ratio of
core radio flux to continuum optical/UV flux that has been proposed
as an "improved orientation indicator," for RLQs. For the details of
the justification of this claim (for $R_{V}$, which is the ratio of
5 GHz flux density of the core to the optical flux density) please
see \citet{wbr95}. The basic idea is that the core radio flux is
from a highly Doppler enhanced relativistic jet and its value is
very sensitive to the line of sight to the jet \citep{lin85}.
Conversely, the optical emission represents the (almost) isotropic
thermal emission from the accretion flow. Comparison of the two is a
crude indicator of the angle that the line of sight makes to the
radio jet \citep{wbr95}. Unfortunately, $R_{V}$ cannot be used
directly since high resolution radio images are not available for
many of the sources in the sample, so the core flux density is not
known. The two point spectral index, $\alpha_{10}^{UV}$, uses high
frequency radio flux as a surrogate for core flux density. Since
large scale radio flux is optically thin, steep spectrum emission
and the unresolved radio core is often flat spectrum, the contribution of the core
emission to the total flux density increases rapidly as the observing frequency
increases. The higher the radio frequency that is used, the more
accurate this surrogate will be. However, since many of the radio
quiet sources had measurements only at 1.4 GHz and 5 GHz, an
extrapolation of the spectrum above 10 GHz in the rest frame is not
justified. The use of the 10 GHz flux density surrogate makes the
interpretation of $\alpha_{10}^{UV}$ as an orientation indicator for
RLQs inferior to $R_{V}$ for SSQs since the lobe flux density can
still dominate the core flux density at 10 GHz in some lobe
dominated quasars with very weak cores. However, in some ways this
broadband spectral index might be considered an improvement on the
orientation indicator, $R_{V}$, since it implements the far UV flux
density rather than the optical flux density, especially when
considering blazars and FSQs. The amount of dilution of the
observed, isotropic, thermal spectrum emitted by the quasar
accretion flow by the steep spectrum, high frequency tail of the jet
synchrotron emission dies off rapidly as the frequency increases.
Thus, for most broad line blazars with a strong nonthermal optical
component, the far UV luminosity is dominated by the thermal
component and is representative of the accretion flow luminosity
\citep{mal86}. For those objects where the thermal component is
small and the synchrotron component is huge, the far UV luminosity
can still be dominated by the nonthermal jet emission, but the
inaccuracy of this estimate of the thermal component is reduced by
an order of magnitude compared to the same estimate in the optical
band.

\par It is desirable to verify the interpretation of
$\alpha_{10}^{UV}$ as an orientation indicator for RLQs. The
fundamental consistency check of interpreting $\alpha_{10}^{UV}$ as
an orientation indicator is that SSQs should have lower values
(weaker cores) than FSQs since they are believed to be viewed with a
line of sight more inclined from the jet axis than FSQs
\citep{ant93,bar89,lin85}. A K-S test indicates that FSQs have
larger values of $\alpha_{10}^{UV}$ (mean of $0.735 \pm 0.142$)
than SSQs (mean of $0.629 \pm 0.106$) at the 99.5\% significance level. Similarly, a Wilcoxon rank
sum test indicates that the ranks of the $\alpha_{10}^{UV}$ values for FSQs
are larger than the ranks of the $\alpha_{10}^{UV}$ values for SSQs at the
99.8\% significance level. These statistical tests lend credence to the
orientation indicator interpretation of $A_{25-80}$ for RLQs.
\par The errors in $A_{25-80}$ in
Figure 1 arise primarily from the uncertainty in $\lambda_{25}$
since the signal to noise ratio is the smallest in the broad wings.
The error in each quantity in equation (1) was individually
estimated and the results were added in quadrature. The error in
$\lambda_{25}$, for example, was achieved by approximating the
region near the 1/4 maximum point of the line profile by a local
polynomial fit. The error in $\lambda_{25}$ was the determined to be
slope of this polynomial ($\partial \lambda / \partial F_{\lambda}$)
at the 1/4 maximum point times the RMS noise level. This naturally
produces larger errors in $A_{25-80}$ for sources with very broad
wings, i.e., the more horizontal the spectrum in the wings, the
larger the slope ($\partial \lambda / \partial F_{\lambda}$) will
be. The error bars in Figure 1 are very conservative estimates of
the 1 sigma errors in the sense that these errors would be greatly
reduced by smoothing the data. The polynomial fit to the noisy data
minimizes residuals, so it should also be close to a fit that would
minimize residuals in smoothed wings as well, but the RMS noise (the
driver of the large errors in Figure 1) would be considerably less.
The errors in $\alpha_{10}^{UV}$ arise primarily from uncertainty in
the 10 GHz flux density from either variability (blazars) or lack of
high frequency data (RQQs).
\par The sample
plotted in Figure 1 includes 28 RQQs, 27 SSQs defined by a spectral
index greater than 0.5 (based on the convention $F_{\nu} \sim
\nu^{-\alpha}$) from 2.7 GHz to 5 GHz (or 1.4 GHz to 5 GHz if 2.7
GHz data was not available) and 40 FSQs (which will be used
interchangeably with the term blazar in the following). The sample
decomposition in Figure 1 also segregates out the blazars that are
known FERMI or EGRET gamma ray loud sources \citep{abd09}. Although
the gamma ray loud subsample of FSQs is too small for meaningful
statistical analysis, it is an intriguing subsample because these
sources tend to have higher apparent superluminal velocities as
measured by VLBI than other FSQs and this has been interpreted as a
more polar line of sight to the jet \citep{lis09,kel04}.
\par As a
check of our measurement technique we note that previously published
$A_{25-80}$ estimates in \citet{wil95,bas05,cor96} for roughly half
the sources in our sample are within the error bars that are shown
in Figure 1. Thus, we do not expect that there are any significant
systematic differences between the measurement techniques employed
here and those of other research teams for the remaining sources in
Figure 1 that have not been previously published. Note that
\citet{mar96,bac04,sul07} subtract out a narrow line component to
compute asymmetry of the broad component, so it is difficult to make
a quantitative comparison to the data presented here.
\begin{figure}
\includegraphics[width=100 mm, angle= -90]{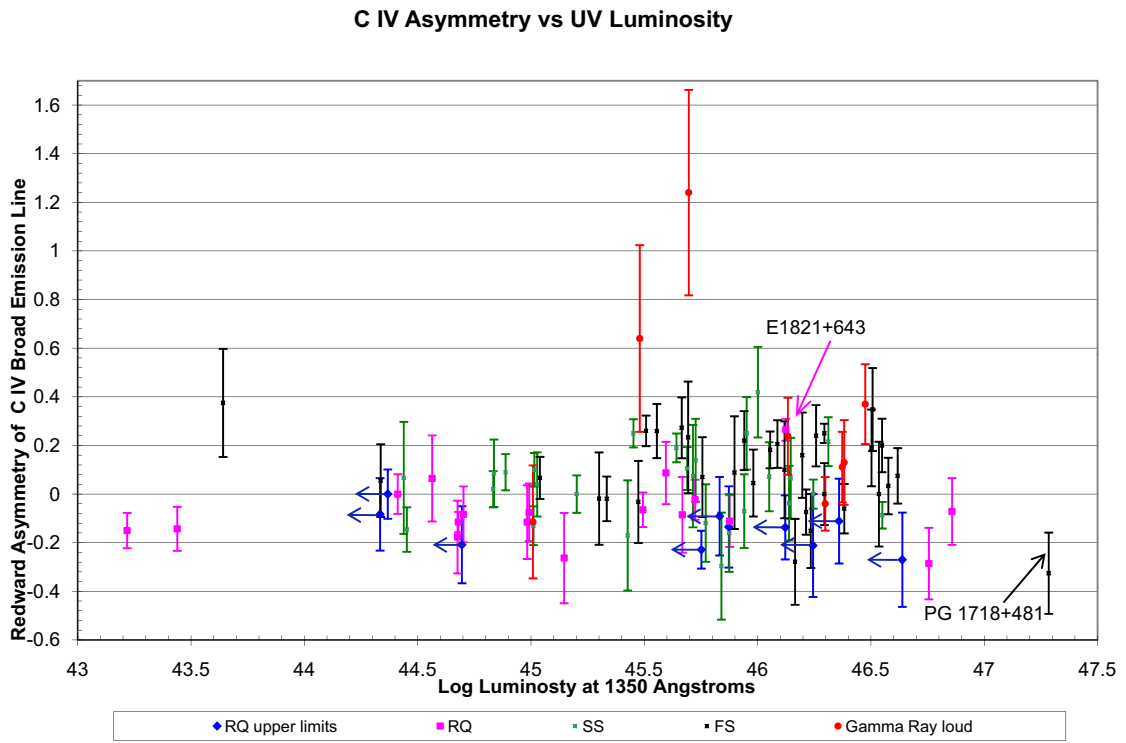}
\caption{Scatter plot of $A_{25-80}$ vs $\lambda L_{\lambda}(1350
\AA)$. In contrast to the results of \citet{cor96}, there is no
correlation between $A_{25-80}$ and the UV luminosity. }
\end{figure}
\begin{table}
\caption{The Probability of a Correlation by Random Chance of
$A_{25-80}$ with $\alpha_{10}^{UV}$}
{\footnotesize\begin{tabular}{ccccc} \tableline \rule{0mm}{3mm}
 Sample & Number & $\alpha_{10}^{UV}$  & 10 GHz Luminosity & $\lambda L_{\lambda}(1350 \AA)$    \\
 \tableline \rule{0mm}{1mm}
  &  & $r_{s}$/P(Null)  & $r_{s}$/P(Null)   &   $r_{s}$/P(null)    \\
 \tableline \rule{0mm}{1mm}
Total      & 95 &  0.575/$<10^{-6}$ & 0.566/$<10^{-6}$ & 0.097/$3.42 \times 10^{-1}$ \\
Radio Loud & 67 &  0.388/$2.20 \times 10^{-3}$ & 0.331/$7.00 \times 10^{-3}$   & 0.017/$8.81 \times 10^{-1}$  \\
Flat spectrum & 40 &  0.376/$1.50 \times 10^{-2}$  &  0.231/$1.47 \times 10^{-1}$  & -0.132/$4.07 \times 10^{-1}$  \\
Steep Spectrum & 27 &  0.145/$4.53 \times 10^{-1}$  &  0.176/$3.68\times 10^{-1}$   & 0.041/$8.26 \times 10^{-1}$  \\
Radio Quiet (All) & 28 &  -0.063/$7.41 \times 10^{-1}$ &  0.001/$9.20 \times 10^{-1}$   & -0.196/$3.12 \times 10^{-1}$  \\
Radio Quiet (No Upper Limits)& 18 &  -0.071/$7.72 \times 10^{-1}$ &  -0.050/$8.41 \times 10^{-1}$   & 0.186/$4.35 \times 10^{-1}$  \\
Radio Quiet (Half Upper Limits)& 28 &  -0.035/$8.49 \times 10^{-1}$ &  0.134/$4.65 \times 10^{-1}$   & -0.183/$3.42 \times 10^{-1}$  \\
\tableline \rule{0mm}{1mm}
\end{tabular}}
\end{table}
\begin{figure}
\includegraphics[width=75 mm, angle= 0]{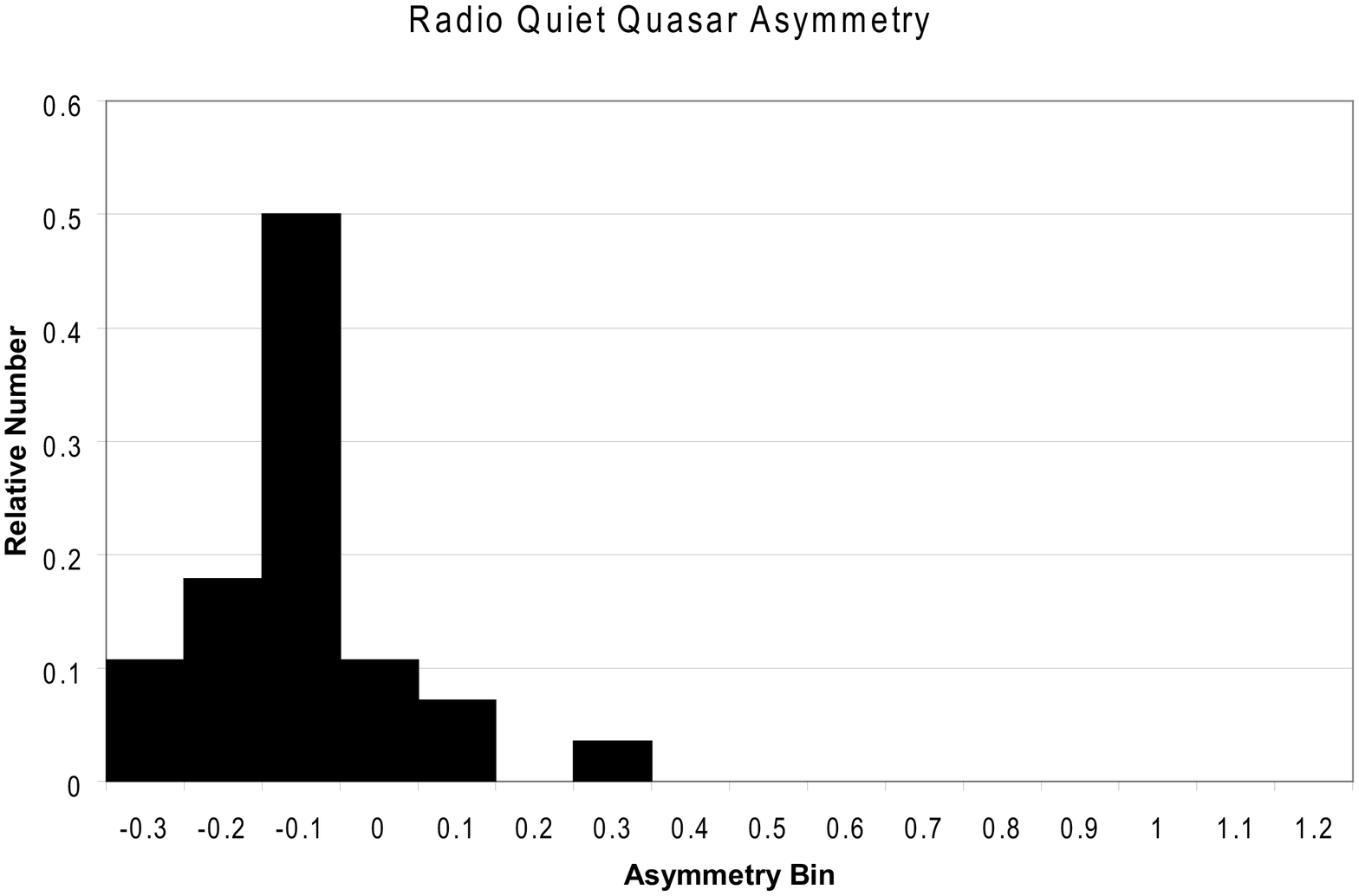}
\includegraphics[width=75 mm, angle= 0]{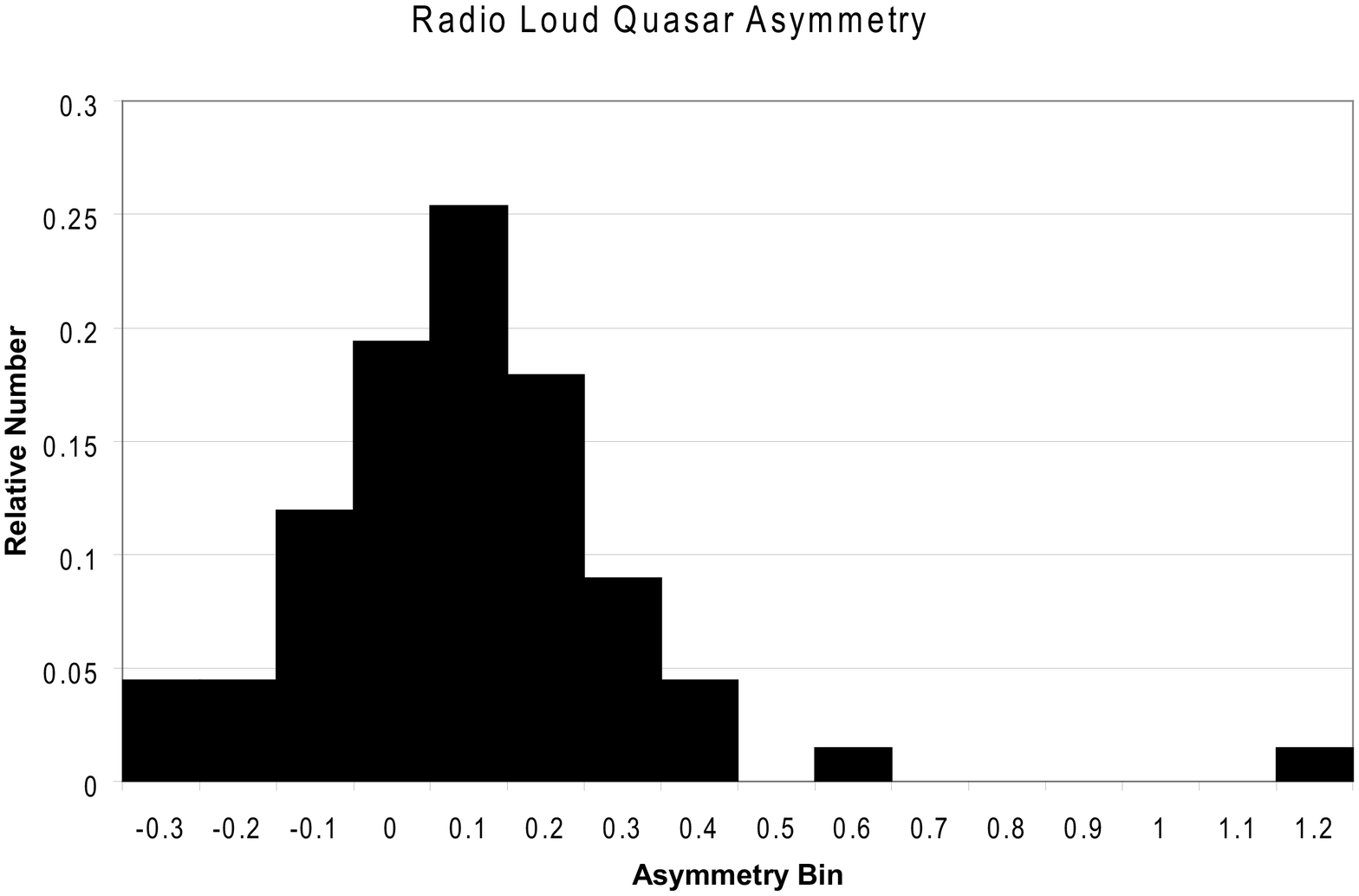}
\includegraphics[width=75 mm, angle= 0]{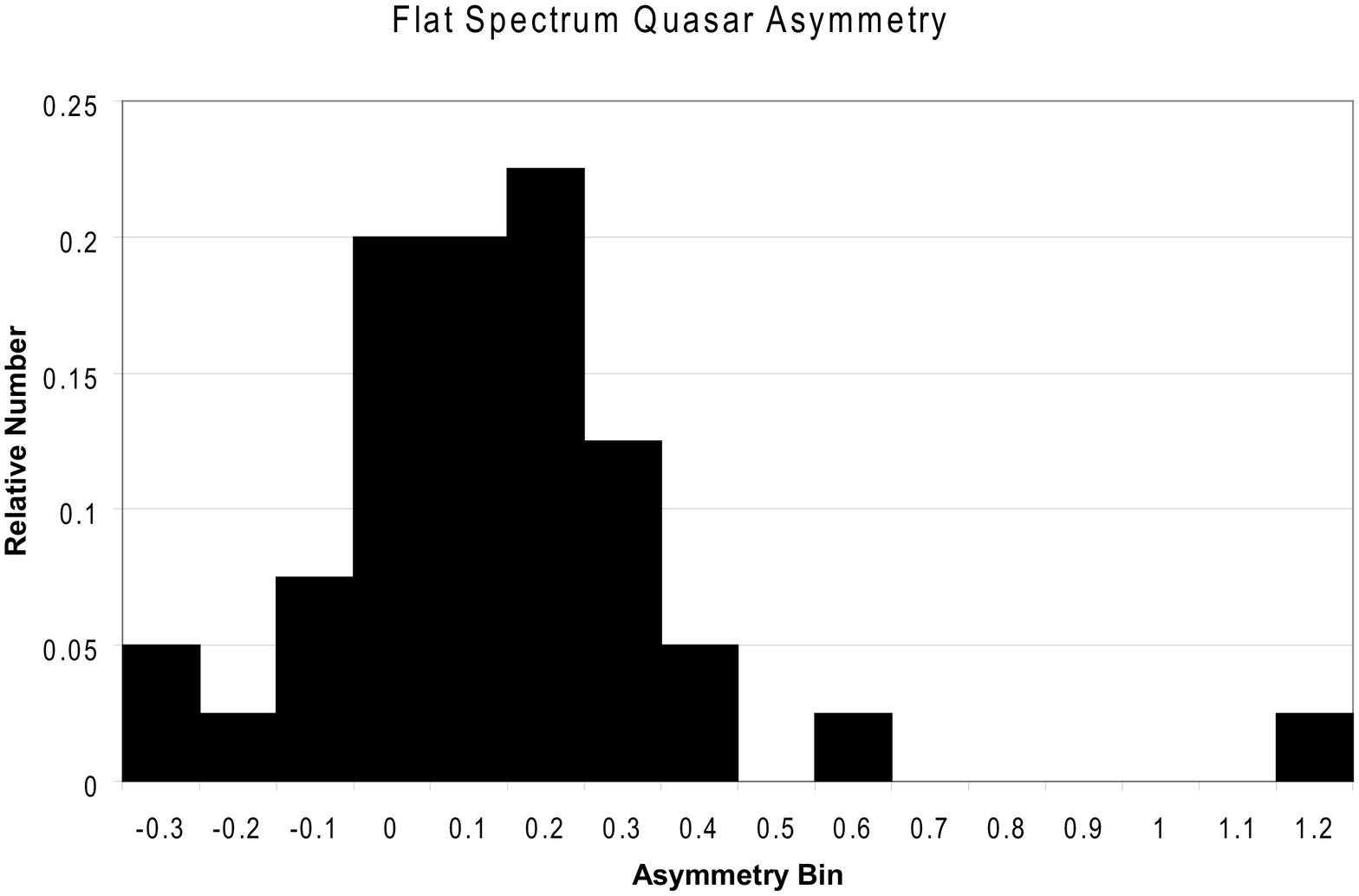}
\hspace{1.1cm}
\includegraphics[width=75 mm, angle= 0]{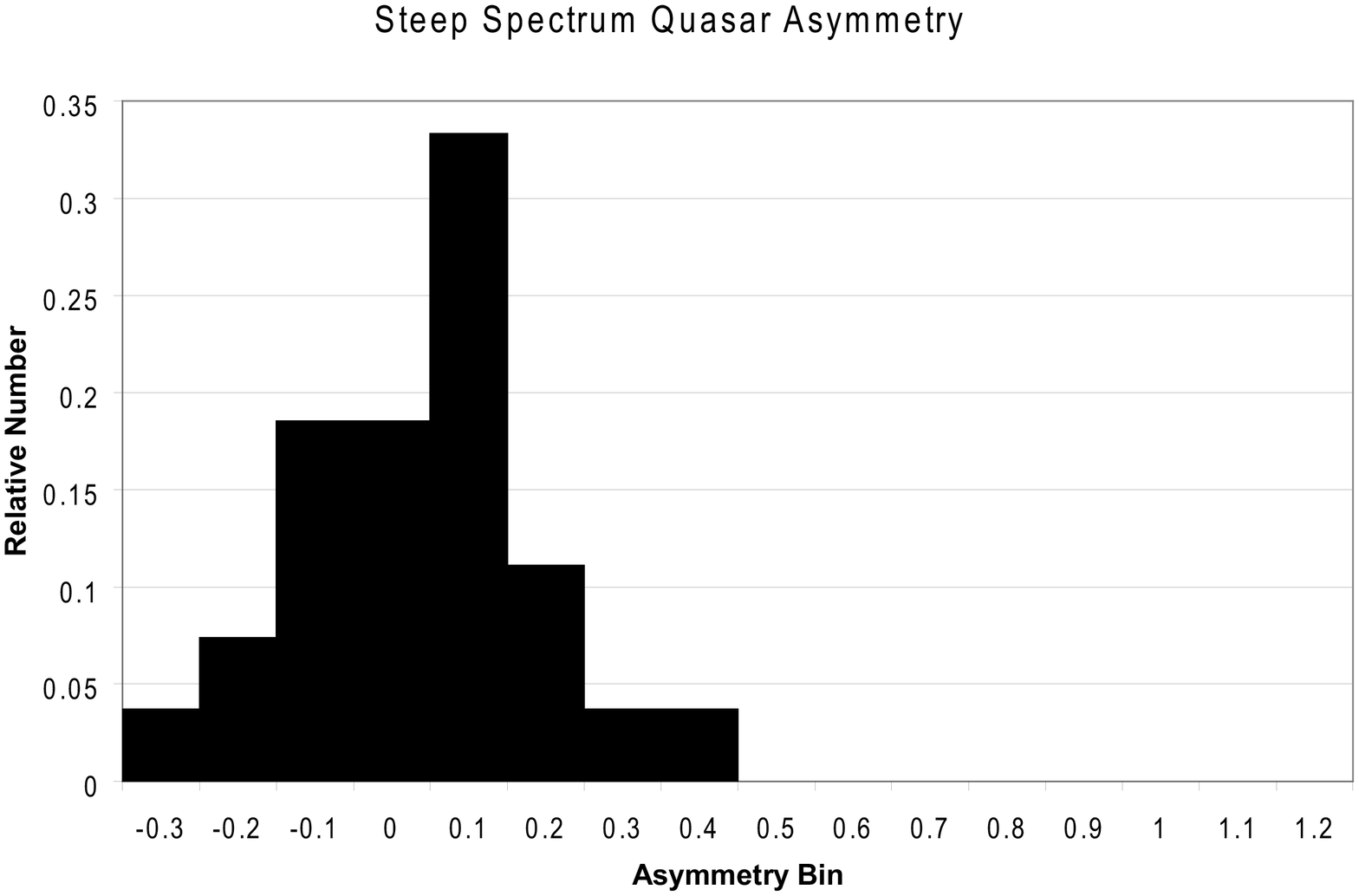}
\caption{Histograms of the normalized distributions of the asymmetry
parameter for each of the subclasses represented in Table 1 }
\end{figure}
\par Figure 2 is a scatter plot of the UV luminosity, $L_{UV} = \lambda L_{\lambda}$ at $1350 \AA$ versus $A_{25-80}$. We do not see any
significant correlation of $A_{25-80}$ and $L_{UV}$ as claimed in
\citet{cor96}. The strength of various correlations (computed by
means of the Spearman rank correlation coefficient) with $A_{25-80}$
are tabulated in Table 1. The Spearman rank correlation coefficient,
$r_{s}$, is listed adjacent to the probability of the null
hypothesis that there is no correlation between the parameters,
labeled "P(Null)." The correlation of $A_{25-80}$ with
$\alpha_{10}^{UV}$ is statistically significant in the total sample
and in the radio loud subsample. The lack of a statistically
significant correlation within the SSQ subsample is likely due to
considerable masking of the core flux density by the often dominant
steep spectrum lobe flux at 10 GHz (eg., P(Null) is reduced to 0.089
if 50 GHz flux density is used as a surrogate instead of 10 GHz).
Since there is a large proportion of upper limits on
$\alpha_{10}^{UV}$ ($\sim$ 1/3) in the RQQ subsample there is no
rigorous way of performing a correlation analysis. In order to deal
with this, the correlation was computed by means of three distinct
models of the data. The first choice in Table 1 was the correlation
with all 28 sources in row 5 using the upper limits on the 10 GHz
flux density as an approximation to the measured values. Secondly,
the uncertainty associated with these upper limits was removed by
just considering the 18 sources without upper limits in row 6. In
row 7, it was assumed that the radio flux density was half of the
upper limit. All three models of the RQQ data showed no evidence of a
correlation of $A_{25-80}$ with other parameters. Thus, it is
concluded that Table 1 indicates no evidence of a statistically
significant Spearman rank correlation within the RQQ subpopulation.
Since the RQQs are randomly clustered in a disjoint region of Figure
1, in the same general direction as the RLQ linear fit (extrapolating toward the
lower left corner), they statistically act as one very heavily
weighted point at the far end of the correlation, thereby enhancing
the statistical significance of the correlation within the total
sample in Table 1.

\section{Interpretation of the Asymmetry}
In the previous section, the correlation between the $A_{25-80}$ and
$\alpha_{10}^{UV}$ was established on a statistical basis in Table
1. There is a vast literature on blueward asymmetries in the CIV in
RQQs. By contrast, this analysis has been directed towards redward
asymmetry. In this section we incorporate previous discussions of
blueward asymmetry into this correlation analysis.
\subsection{The Distribution is Bimodal}
It is shown that there are two disjoint pieces of information
contained within the scatter of the data in Figure 1. First, we consider
the distributions of asymmetry in Figure 3 for each quasar subclass
represented in Table 1. There is a clear distinction in the
asymmetry of the RQQs with that of the RLQs as well as with the
radio loud sub-populations FSQ and SSQ, individually, as evidenced
by a K-S test. The maximum difference D and the K-S probability that
the pairs of samples are drawn from the same population are listed
in Table 2.
\begin{table}
\caption{The Probability of $A_{25-80}$ Being Drawn From the Same
Population} \begin{tabular}{cccc} \tableline \rule{0mm}{3mm}
 Sample 1/Number & Sample 2/Number & Maximum Difference, D & Probability   \\
\tableline \rule{0mm}{1mm}
RQQ/28 & RLQ/67 & 0.5917 & $< 0.001$   \\
RQQ/28 & FSQ/40 &  0.6607 & $< 0.001$     \\
RQQ/28 & SSQ/27 &  0.4907  &  0.002  \\
FSQ/40  & SSQ/27 &  0.2718  &  0.138    \\

\tableline \rule{0mm}{1mm}
\end{tabular}
\end{table}
The K-S test results in Table 2 indicate that $A_{25-80}$ is a
measurable difference between RLQs and RQQs with high statistical
significance. This bimodality is mirrored in the radio sector, these
sources are bimodal in the power of the radio jet. The bimodality is
manifest in Figure 1 as RQQs have $\alpha_{10}^{UV}< 0.3$ and RLQs
have $\alpha_{10}^{UV}> 0.3$.
\subsection{Redward Asymmetry is Associated with
a Strong Radio Jet} The Spearman rank correlation for the RQQs in
Table 1 strongly supports the notion that there is no correlation
between $A_{25-80}$ and other parameters. This might be expected
since \citet{bas05} found that the asymmetry did not correlate with
any other measured property in a sample of predominantly radio quiet
quasars. The lack of signs of correlation in the RQQs and the strong
correlation of the RLQs with $\alpha_{10}^{UV}$ in Table 1 is
another aspect of the bimodality of the two populations noted in the
previous subsection. In terms of the correlation in Figure 1, the
RQQs are essentially objects with virtually no 10 GHz flux density
to first order and they are therefore clustered in the far left hand
bottom corner of the scatter plot. This clustering in the lower left
hand bottom corner is most naturally explained by the fact that
\textbf{redward asymmetry in the CIV BEL profile is associated with the presence of a radio jet}, a
feature that is weak in RQQs by definition.
\subsection{Blueward Asymmetry is Associated with the Accretion Disk}Consider the
vast literature demonstrating a blueward asymmetry in RQQ CIV BELs
\citep{wil84,bro94,mar96,bas05,sul07}. Our data in Figure 3 agrees
with this notion, the distribution of $A_{25-80}$ for RQQ is heavily
skewed towards negative values. The physics of the radio quiet
quasar phenomenon typically induces a blueward asymmetry (through an
unknown mechanism). The physics creating radio quiet quasars is
generally believed to be the thermal luminosity resulting from
dissipation in an accretion flow onto a supermassive black hole
\citep{sun89}. Furthermore, in this standard model, the BELs
represent reprocessed thermal emission in gas that is photo-ionized
by the accretion flow emissivity. It follows that the blueward
asymmetry of the CIV BELs is empirically determined to arise in the
combined dynamical system of the accretion flow, the resultant
radiation field and the broad emission line gas. It has been
suggested that the blue excess in the CIV BEL is evidence of an
outflowing wind from the accretion disk in RQQs \citep{mar06}.
\par Now consider the dynamical environment of RLQs, for which there are two central
engines. The first one is associated with the accretion flow and its
powerful thermal radiation, i.e., the same as the RQQs. In addition
there is a second central engine with similar power that drives a
radio jet. The accretion disk is associated with an excess of CIV
BEL gas that is blueshifted relative to the CIV peak. Based on the
correlations in Table 1 and Figure 1, the radio jet central engine
is associated with an excess of CIV BEL gas that is redshifted
relative to the CIV peak. A RLQ has both central engines and there
are competing effects. The first is related to the accretion disk
and has a propensity to be associated excess blueshifted emission in
the CIV BEL. The other is related to the radio jet and is associated
with excess redshifted emission in the CIV BEL. This has been
demonstrated empirically, so we don't know the exact mechanism which
causes the accretion disk (radio jet) to produce blueshifted
(redshifted) CIV emission. These are competing effects that can show
up in different relative strengths in principle. The statistical
evidence in Table 1 that there are competing affects is that
$A_{25-80}$ is more strongly correlated with the logarithmic ratio
of the radio luminosity (loosely associated with jet power) to the
UV luminosity (accretion induced thermal luminosity) than to either
of the luminosities (UV or 10 GHz) in RLQs. It is the relative
strength of the jet to the accretion disk, not just the strength of
the jet that produces the strongest correlation. Physically
speaking, each central engine can be extant with a wide spread of
strength independent of the other central engine's strength and by
some mechanism each induces broad wing emission in the BEL gas that
exists in a myriad of possible enveloping environments - this is
undoubtedly a complicated dynamical system. This is very different
than what is the case for RQQs. The extra degree of freedom for
creating emission in the CIV broad wings should result in
\textbf{both blue asymmetries and red asymmetries in RLQ CIV BELs}
and \textbf{much larger cosmic scatter in the RLQ asymmetry
properties than for RQQs}. Evidence for the first of these
conditions is indicated by the distribution of $A_{25-80}$ in the
RLQ histogram in Figure 3. The second condition is supported by the
comparison of the histograms for RQQs and RLQs in Figure 3, the RLQ
distribution is much wider. The scatter of the data is consistent
with two competing sources of BEL wing gas.
\par We can test this idea of two competing effects by looking at
the two outliers in detail. The FSQ PG 1718+481 with $A=-0.33$ and
the RQQ E1821+643 with $A=0.26$. If the interpretation given is
correct, then there should be evidence that PG 1718+481 has strong
accretion disk radiation and E1821+643 has a powerful jet. The
primary indicator that a luminous accretion flow (a quasar) is
present in a galactic nucleus is the detection of a large UV peak in
the spectral energy distribution \citep{sun89}. Thus, this first
prediction seems to be verified by the fact that $\lambda
L_{\lambda}(1350\AA)= 2\times 10^{47}\mathrm{ergs/s}$ for PG
1718+481, by far the largest value in Figure 2. This large UV flux
accounts for the small value of $\alpha_{10}^{UV} =0.424$, one of
the lowest in the RLQ sample. The continuum flux is almost certainly
from the accretion disk and not the high frequency tail of the
synchrotron emission from the jet based on the following four facts.
First, the spectral index is too flat from $3350 \AA$ to $1650 \AA$,
$\alpha_{\nu} =0.76$ to be a blazar synchrotron emission and the EWs
of the BELs are too large for the continuum to be dominated by
synchrotron emission \citep{sti91,wil95}. Also, the NASA
Extragalactic Database notes that it is not optically variable and
the optical polarization is low, both inconsistent with a
synchrotron interpretation of the continuum. Within the two
component model, the accretion disk luminosity is so strong the it
induces (through a yet to be determined physical process) a large
blueward excess which swamps any redwing contributions induced by
the relativistic jet (through a yet to be determined process).
\par The case for E1821+643 is not as strong, but still highly
compelling. This quasar is a powerful radio source with a 151 MHz
luminosity intermediate between that of a very strong FR I and an FR
II radio source \citep{blu01}. Physically, the 151 MHz luminosity is
believed to be a measure of the long term time averaged kinetic
power delivered by the jet to the distant radio lobes \citep{wil99}.
The methods of \citet{wil99} indicate a jet kinetic luminosity of
$\approx 2 \times 10^{44}$ ergs/s. VLA observations indicate that
the source is core dominated and flat spectrum \citep{blu96}. The
core spectrum is highly inverted between 5 GHz and 15 GHz with
$\alpha = -0.94$ with a 15 GHz flux density of 25 mJy
\citep{blu95,blu96}. The peak of the spectral energy distribution of
the core is unknown. VLBA observations show that the core size is
less than 1 pc which bounds the brightness temperature $>1.4 \times
10^{9}\, ^{\circ}K$ \citep{blu96}. The flat spectrum nature of the
jet and its large high frequency luminosity are consistent with a
powerful jet viewed close to the line of sight. The very large high
frequency flux of the core might be indicative of a current episode
of increased jet power. Within the two component model described
above, The dominant flat spectrum core is the manifestation of a
powerful FR II jet viewed near the line of sight. This dynamics and
geometry of the current episode of jet ejection are inducing a
redward asymmetry that swamps the blueward asymmetry produced by the
powerful accretion disk.
\par It is interesting that the outlier, E1821+643, in the scatter plot in Figure 1
is also a well studied outlier in the quasar literature. Despite the
radio characteristics just described it is formally radio quiet
(with $R \approx 1.5$) per the standard definition of radio loudness
in the footnote on the first page of this Letter \citep{blu96}. The
reason that it is always classified as radio quiet in spite of a
powerful radio jet is that the optical/UV luminosity is extremely
large, $L_{UV} = 1.3 \times 10^{46} \mathrm{ergs/s}$ (see Figure 2),
making it one of the most luminous quasars at $z < 0.5$
\citep{lac99}. This circumstance results in E1821+643 lying within
the range (at the high end) of $\alpha_{10}^{UV}$ of the RQQs in
Figure 1, $\alpha_{10}^{UV} = 0.200$, even though it has a powerful
jet. It is a rare case of a quasar that lies in a giant elliptical
galaxy at the center of a cluster, but is not a RLQ \citep{lac92}.
This object has observational properties that are typical of both
RQQs and RLQs, hence it has properties typical of both RLQs (large
$A_{25-80}$) and RQQs (small $\alpha_{10}^{UV}$) in the scatter plot
in Figure 1. A major advantage of the use of $\alpha_{10}^{UV}$ in
the analysis in this Letter is that is independent of the
classification of a quasar as a RLQ or a RQQ.

\section{Discussion}In this letter, we analyzed the line asymmetries in
 a large sample of quasar CIV BELs.
 \newcounter{bean}
 The following results were shown
 to be statistically significant.
 \begin{list}
 {A--\Roman{bean}}{\usecounter{bean}
 \setlength{\rightmargin}{\leftmargin}}
 \item $A_{25-80}$ is correlated with $\alpha_{10}^{UV}$ in the total QSO
 sample, the RLQ subsample and the FSQ subsample (see Figure 1 and Table 1).
 \item $A_{25-80}$ is not correlated $\alpha_{10}^{UV}$ in the RQQ subsample
 (see Figure 1 and Table 1).
 \item The distribution of $A_{25-80}$ is bimodally split in the radio sector
 of QSO parameter space: RQQs have smaller, usually negative $A_{25-80}$ values
(i.e., RQQS generally have blue asymmetric CIV BEL profiles), in
contrast to the RLQs that have $A_{25-80}$ values that are more
broadly distributed with a tendency toward red asymmetric CIV BEL
profiles (see Figures 1 and 3). Furthermore, the RQQ $A_{25-80}$
values are not correlated with $\alpha_{10}^{UV}$ while the RLQ
$A_{25-80}$ values are correlated with $\alpha_{10}^{UV}$ (see Table
1).
\end{list}

These results are mathematically robust.
\newcounter{numb}
More speculatively we physically interpret these facts as arising
from a two component model with the following properties (with the
justifying information in parenthesis).

\begin{list}
{B--\Roman{numb}}{\usecounter{numb}
 \setlength{\rightmargin}{\leftmargin}}
\item The blue asymmetry in the CIV BEL, $A_{25-80}<0$, is associated with accretion disk physics and its
interaction with the enveloping gaseous environment (The RQQ
histogram in Figure 3 and the standard RQQ model \citep{sun89}).
 \item The red asymmetry in the CIV BEL, $A_{25-80}>0$, is
associated with the central engine of a powerful relativistic jet
and its interaction with the enveloping gaseous environment (The
majority of RLQ in the histogram in Figure 3 have $A_{25-80}>0$, almost no
RQQ have $A_{25-80}>0$ in the histogram. The largest $A_{25-80}>0$ RQQ source has a
powerful jet, see section 3.3).
\item The
tendency for the radio jet to be associated with red asymmetry in
the CIV BEL appears to increase as the line of sight gets closer to
the radio jet axis (The correlation of $A_{25-80}$ with
$\alpha_{10}^{UV}$ in Table 1 for RLQs and FSQs).
\item These
competing effects of the accretion disk and the jet can exist in
various ratios within in RLQs and give rise to a panoply of CIV lines
shapes that are described by the correlation in Figure 1 (The RLQ
histogram in Figure 3).
\end{list}

 \par Finally, the results above are cast in the
light of the 9 historical findings that were noted in the
Introduction. The result A-I above (the correlation in Figure 1),
incorporates the findings of 2 (CDQS have larger $A_{25-80}$ than
other RLQs), 4 (the largest $A_{25-80}$ are in RLQs), 5 (RLQs have
redder CIV BELs than RQQs), 6 (composite CIV has redder wings for
RLQs than RQQs) and 7 (some superluminal quasars have large red wing
excess) from the Introduction. The finding 8 (RLQs have a larger
spread in $A_{25-80}$ than RQQs) from the Introduction is verified
by the relative spread in the histograms in Figure 3 and is
explained by B-IV of the consequences of the physical interpretation
that are noted above. Historical finding 9 (RQQS tend to have $A<0$)
is consistently described by results A-II and A-III noted at the
beginning of this section. Figure 2 shows that the conclusion of
study 3 ($A_{25-80}$ correlated with $L_{UV}$) was an artifact of a
sample in which the most luminous sources were often RLQs, as
suggested in \citet{cor97}. There is no support for the conclusion
found in study 1 (SSQs have larger $A_{25-80}$ than FSQs) in
agreement with \citet{wbr96} who note that the result in study 1 was
only marginally significant.

\begin{acknowledgements}
I would like to thank Matt Malkan for sharing his expertise and encouraging me to develop my
preliminary results and publishing them. I would also like to thank an anonymous referee who offered
many ideas that improved the manuscript.
\end{acknowledgements}

\end{document}